\documentclass[10pt,letterpaper]{article}

\usepackage{cogsci}
\usepackage{pslatex}
\usepackage{apacite}
\usepackage{amsmath}
\usepackage{graphicx}

\title{Outgroup Homogeneity Bias Causes Ingroup Favoritism}
 
\author{{\large \bf Marcel Montrey (marcel.montrey@mail.mcgill.ca)} \\
  Department of Psychology, McGill University\\
  2001 McGill College Avenue, Montreal, QC H3A 1G1 Canada
  \AND {\large \bf Thomas R. Shultz (thomas.shultz@mcgill.ca)} \\
  Department of Psychology and School of Computer Science, McGill University \\
  2001 McGill College Avenue, Montreal, QC H3A 1G1 Canada}

\begin{document}

\maketitle

\begin{abstract}
Ingroup favoritism, the tendency to favor ingroup over outgroup, is often explained as a product of intergroup conflict, or correlations between group tags and behavior.
Such accounts assume that group membership is meaningful, whereas human data show that ingroup favoritism occurs even when it confers no advantage and groups are transparently arbitrary.
Another possibility is that ingroup favoritism arises due to perceptual biases like outgroup homogeneity, the tendency for humans to have greater difficulty distinguishing outgroup members than ingroup ones.
We present a prisoner's dilemma model, where individuals use Bayesian inference to learn how likely others are to cooperate, and then act rationally to maximize expected utility.
We show that, when such individuals exhibit outgroup homogeneity bias, ingroup favoritism between arbitrary groups arises through direct reciprocity.
However, this outcome may be mitigated by:
(1) raising the benefits of cooperation,
(2) increasing population diversity, and
(3) imposing a more restrictive social structure.

\textbf{Keywords:} 
ingroup favoritism;
outgroup homogeneity;
direct reciprocity;
Bayesian learning;
conditional expected utility
\end{abstract}

\section{Introduction}
Ingroup favoritism is the tendency for people to favor members of their own group over members of other groups.
It manifests as a bias in how people evaluate others~\cite{Brewer1979,Galinsky2000}, distribute rewards~\cite{Tajfel1971}, mete out punishments~\cite{Bernhard2006}, and decide whether or not to cooperate~\cite{Dorrough2015}.
Though readily elicited in both natural~\cite{Rand2009} and arbitrary groups~\cite{EffersonLalive2008,Galinsky2000}, the existence of ingroup favoritism is puzzling.
It often neither improves the population's average outcome, nor maximizes that of the individual~\cite{Nakamura2012}.
Disagreement even exists as to whether ingroup favoritism is better understood as a preference for improving the welfare of ingroup over outgroup, or as a product of divergent beliefs about how these groups behave~\cite{Everett2015}.
However, empirical work suggests that people generally expect ingroup members to act in a cooperative manner~\cite{Brewer2008,Yamagishi1999}, and meta-analysis confirms that this expectation is indeed stronger toward ingroup than outgroup~\cite{Balliet2014}.
A promising avenue for explaining ingroup favoritism therefore seems to be understanding how people arrive at these beliefs.
In short, why are ingroup members seen as more cooperative than outgroup ones?

Many theoretical models have addressed this question.
One common approach is to assign phenotypic tags to individuals, and then see what is required to elicit ingroup favoritism.
Such models have shown that ingroup favoritism may be selected for when tags are not arbitrary, but rather correlate with behavioral traits~\cite{Jansen2006,Masuda2007,Traulsen2008}.
These traits typically include willingness to cooperate, or suitability as a cooperative partner.
Ingroup favoritism may thus occur when tags convey information useful in guiding the individual's own actions.
Other models explain ingroup favoritism as a product of intergroup conflict~\cite{Choi2007,Garcia2011,Konrad2012}, where group membership may be arbitrarily decided, but remains relevant from a competitive point of view.
However, a classic empirical finding is that humans show ingroup favoritism even when groups are both explicitly arbitrary and functionally irrelevant~\cite{Billig1973,Locksley1980}.
So why should ingroup favoritism occur even when group membership is meaningless, and such outcomes are maladaptive?

One explanation is that ingroup favoritism may arise through cognitive or perceptual limitations~\cite{Masuda2012}.
For instance, humans are known to perceive outgroup members as more similar to one another than ingroup members, a bias known as outgroup homogeneity~\cite{Judd1988}.
By approximating individuals' characteristics through a single group stereotype, this may serve to reduce cognitive burden~\cite{Masuda2012}.
\citeA{Masuda2012} studied the implications of such a bias on indirect reciprocity, where cooperation is conditioned on whether or not partners maintain a good reputation.
In the simplest such scheme, an individual's reputation improves when it is observed to cooperate, and suffers when it is observed to defect;
in more complicated schemes, reputation may, for example, also be gained by being observed punishing a defector, or lost by being observed cooperating with one.
To simulate outgroup homogeneity, individuals were allowed to observe accurate reputation information about ingroup members, but only group-level information about outgroup members.
Ingroup favoritism occurred, but only when additional assumptions were invoked, such as individuals using a different rule for attributing reputation to ingroup than to outgroup members.
A follow-up model by \citeA{Nakamura2012} eliminated the need for such double standards, and also produced ingroup favoritism through indirect reciprocity.
However, this time the result was contingent on reputation information being only shareable within groups, but not between them.

Here, we show that complex rules for assigning and sharing reputation are not needed to explain ingroup favoritism between arbitrary groups.
Rather, outgroup homogeneity bias may drive ingroup favoritism through a much simpler mechanism: direct reciprocity (learning through personal experience).
We create an agent-based computational model, where individuals are assigned arbitrary group tags, and then play a prisoner's dilemma~(PD) game.
These individuals use Bayesian inference to learn how likely others are to cooperate or defect, and then act rationally by maximizing their conditional expected utility.
We show that introducing outgroup homogeneity bias into this minimal setting is sufficient to produce strong ingroup favoritism, and propose several ways of mitigating this outcome.

\section{Model}
\subsection{Prisoner's Dilemma}
We consider a PD game where pairs of neighboring individuals interact by either cooperating (C) or defecting (D).
The game is parameterized by two values:
the benefit of receiving cooperation, $b$, and the cost of cooperating, $c$.
When both individuals cooperate, both receive the benefit of cooperation, but pay the cost of cooperating, $b-c$.
If one individual defects while the other cooperates, then the cooperator pays the cost while receiving no benefit, $-c$, while the defector pays no cost but receives the full benefit, $b$.
When both individuals defect, neither receives the benefit nor pays the cost.
The following table summarizes the row player's payoffs:
\begin{center}
\begin{tabular}{ r | c c }
	  & C & D \\ \hline
	C & $b-c$ & $-c$ \\
	D & $b$ & $0$ \\
\end{tabular}
\end{center}
As long as $b > c > 0$, each player's payoff is always improved by defecting, no matter what the other player does.
This makes the game a dilemma, because although the best individual outcome is unilateral defection, the best average outcome is mutual cooperation.

\subsection{Social Structure}
In PD, ingroup favoritism is operationalized as a higher rate of cooperation toward ingroup partners than outgroup ones~\cite{Dorrough2015,Fu2012,Gray2014,Masuda2012}.
For ingroup favoritism to be possible, cooperation must also be possible.
By constraining which individuals interact, we promote repeat interactions, which in turn promotes cooperation~\cite{Szabo2007}.
For each run, we generate a random $r$-regular graph~\cite{Bollobas2001} with 1000 vertices, using Steger and Wormald's~\citeyear{Steger1999} algorithm.
Each vertex represents an individual, and each edge represents a connection between neighbors.
This graph governs interactions by limiting individuals to playing PD exclusively with their neighbors.

\subsection{Group Tags}
Individuals are divided into $m$ groups, where group membership is represented by a tag visible to all other individuals.
By default, $m=2$, though it may take other values, as long as $m>1$.
Otherwise, tags cease to represent group membership, and instead become a universally shared characteristic.
Each individual is randomly assigned a tag, such that each group has the same initial number of members.
When replacement occurs, newcomers are assigned a tag uniformly at random.

\subsection{Rational Bayesian Learning}
Learning involves estimating a pair of parameters for each partner $i$ that the individual interacts with.
The first parameter $p_i$ represents the estimated probability that partner $i$ will cooperate with the individual, given that the individual cooperates with that partner, $Pr(C_i|C)$.
The second parameter $q_i$ estimates the probability that partner $i$ will cooperate, given that the individual defects against that partner, $Pr(C_i|D)$.
Because the game is simultaneous, actions cannot be conditioned on those of the partner.
However, there is no \textit{a priori} reason for individuals to know this, and indeed repeated interactions cause $p_i$ and $q_i$ to diverge, as individuals change their behavior in response to that of their partner.
Individuals use Bayesian inference to arrive at point estimates for $p_i$ and $q_i$.
Here, the posterior predictive distribution corresponds to the posterior mean~\cite{Griffiths2008},
\begin{equation}
p_i := \frac{n_{CC} + \alpha + 1}{n_{CC} + n_{CD} + \alpha + \beta + 2} \quad q_i := \frac{n_{DC} + \alpha + 1}{n_{DC} + n_{DD} + \alpha + \beta + 2},
\label{learning-rule}
\end{equation}
where $n_{AB}$ counts the number of times the individual took action $A$ when partner $i$ took action $B$.
Similarly, $\alpha$ and $\beta$ are pseudocounts~\cite{Griffiths2008} that encode prior knowledge or expectations about the frequency of cooperation and defection, respectively.
These take the value $\alpha = \beta = 0$, which represents a neutral prior (uniform distribution), where neither cooperation nor defection is seen as inherently more likely.

By default, individuals maintain a pair of $p$ and $q$ values for each partner $i$.
However, individuals exhibiting outgroup homogeneity bias do not distinguish between outgroup members, so they instead track a single pair of values, $p_j$ and $q_j$, for each outgroup $j$.
Outgroup homogeneity thus causes individuals to treat outgroups as if they were a single individual.

Individuals act rationally on their Bayesian estimates, so as to maximize their conditional expected utility~\cite{Jeffrey1990}.
More formally, an individual cooperates if
\begin{equation}
	p b - c > q b,
    \label{decision-rule}
\end{equation}
and defects otherwise.
To give individuals a chance to sample both actions, we implement a small trembling-hand parameter~\cite{Selten1975}.
When an individual selects an action, with a small probability $\epsilon = 0.01$, it takes the opposite action instead.
Removing this parameter (setting $\epsilon=0$) does not qualitatively alter our results.

\subsection{Simulation}
At each time step, individuals interact with their neighbors in random order.
Interactions involve selecting an action (C or D), and then playing PD.
After each interaction, individuals note the outcome of the game, and then update their estimates $p$ and $q$.
Once everyone has finished playing, individuals are subjected to a $0.01$ probability of being replaced.
Newcomers are assigned a group tag uniformly at random, and have no knowledge of their predecessor's $p$ and $q$ values.
Because there is no selection over genotypes, ingroup favoritism cannot evolve, but arises through phenotypic plasticity (i.e. learning) instead.
We run simulations for 1000 time steps, by which time cooperation rates have long stabilized.
All results are averaged across 20 independent runs, and stabilized cooperation rates are further averaged over the last 100 time steps.
In all figures, line width represents 95\% confidence intervals.

\section{Results}

\begin{figure}[t]
  \begin{center}
  \includegraphics[width=\linewidth]{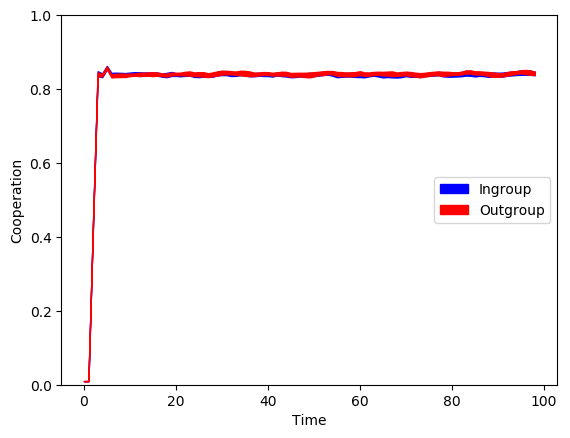}
  \end{center}
  \caption{Ingroup and outgroup cooperation rates over time, in the absence of outgroup homogeneity bias. Cooperation rates climb rapidly as individuals learn that defection is met with defection. Ingroup cooperation rates mirror outgroup cooperation rates, because group membership is irrelevant.} 
  \label{unbiased}
\end{figure}

We first consider unbiased individuals, connected to $r=10$ random neighbors, where the benefit of receiving cooperation ($b=3$) moderately exceeds the cost of giving it ($c=1$).
In the first few time steps, cooperation rates are near-zero (Figure~\ref{unbiased}).
Recall that na\"ive agents have a uniform prior, meaning that cooperation and defection are seen as equally likely.
However, the expected utility of unilateral defection is higher than that of mutual cooperation (Inequality~\ref{decision-rule}), and so virtually everyone defects, hoping to take advantage of a cooperating partner.
Individuals quickly learn that defection is met with defection, which lowers their estimate of $q$.
With unilateral cooperation seeming increasingly unlikely, $qb$ falls below $pb-c$, and individuals seek out mutual cooperation instead.
As cooperation is met with cooperation, estimates of $p$ increase, and high rates of cooperation ($\sim 84\%$) are established.
Although individuals are assigned to $m=2$ random groups, group membership is irrelevant, and so ingroup and outgroup cooperation rates do not differ.

\begin{figure}[t]
  \begin{center}
  \includegraphics[width=\linewidth]{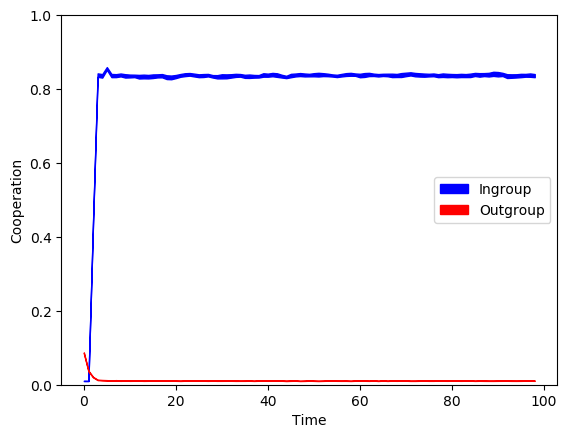}
  \end{center}
  \caption{Ingroup and outgroup cooperation rates over time, with outgroup homogeneity bias. Here, individuals track outgroup members' behavior at a group rather than individual level. Breakdowns in cooperation result in cascades of defection involving entire groups, rather than just the offending individual, resulting in strong ingroup favoritism.} 
  \label{outgroup-homogeneity}
\end{figure}

When outgroup homogeneity bias is introduced, outgroup members cease being treated as individuals, but as representatives of their group.
By generalizing the outcome of each interaction to other members of the outgroup, individuals learn that defection does not yield cooperation even more rapidly than when playing against ingroup members.
This causes an initial spike in outgroup cooperation (Figure~\ref{outgroup-homogeneity}).
However, although ingroup cooperation is a bit slower to get going, it alone persists.
To understand why, consider what happens when an individual cooperates, but its partner defects.
If partner $i$ is an ingroup member, then the individual revises its beliefs about that partner's willingness to cooperate, and $p_i$ declines.
Soon, $p b - c$ drops below $q b$, and the individual ceases to cooperate.
Once partner $i$ learns that defection does not evoke cooperation, its $q$ falls low enough for it to also seek mutual cooperation.
Any successful instance of mutual cooperation promotes further cooperation, causing $p$ values to increase, entrenching that behavior.
However, if partner $i$ is an outgroup member, then the individual does not know who to blame for the partner's unilateral defection.
The individual thus revises its beliefs about the entire group's willingness to cooperate, and $p_j$ declines.
This causes the individual to punish not just the defecting partner, but also any others from that group.
Those neighbors then punish the focal individual, as well as members of its group, for this seemingly unprovoked hostility.
Intergroup defections thus bring about not just punishment of the offending individual, but also a cascade of retributive defections.
Outgroup cooperation is prohibitively difficult to establish and maintain under such conditions, resulting in strong ingroup favoritism.

\begin{figure}[t]
  \begin{center}
  \includegraphics[width=\linewidth]{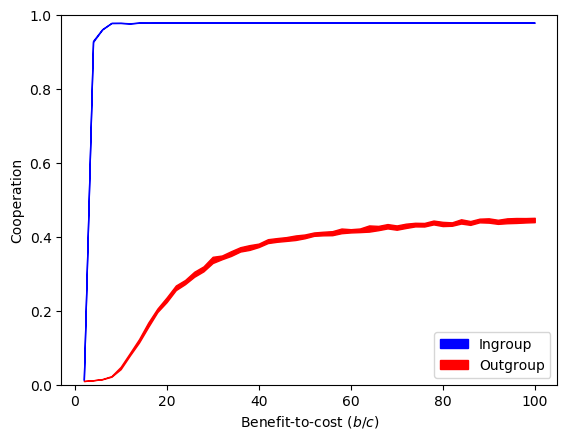}
  \end{center}
  \caption{Stabilized ingroup and outgroup cooperation rates for various benefit-to-cost ($b/c$) ratios. Increasing the $b/c$ ratio favors cooperation more broadly by making the temptation to defect less appealing, thus reducing ingroup favoritism.} 
  \label{bc-ratio}
\end{figure}

We next consider various parameters that may mitigate this outcome.
For example, increasing the trembling-hand parameter $\epsilon$ reduces ingroup favoritism, albeit in a somewhat trivial manner.
The more errors individuals commit in taking their desired action, the more this increases (unwanted) outgroup cooperation and decreases (desirable) ingroup cooperation.
Such effects offer relatively little additional insight, however, because ingroup favoritism is merely harder to enact, rather than less sought after.

Of greater theoretical interest is the effect of increasing the benefit-to-cost ratio of cooperation.
Doing so raises both ingroup and outgroup cooperation, which in turn reduces ingroup favoritism (Figure~\ref{bc-ratio}).
Higher $b/c$ ratios represent more cooperative games, where mutual cooperation is more rewarding, and the temptation to defect is reduced (i.e. Inequality~\ref{decision-rule} becomes primarily driven by $p$ and $q$ values, rather than $c$).
Whereas the ingroup cooperation rate rapidly approaches a ceiling, the outgroup cooperation rate has more room to grow.

\begin{figure}[t]
  \begin{center}
  \includegraphics[width=\linewidth]{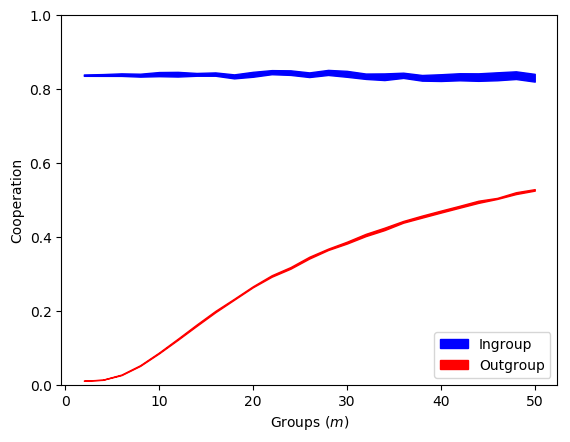}
  \end{center}
  \caption{Stabilized ingroup and outgroup cooperation rates for various numbers of groups ($m$). Increasing population diversity reduces ingroup favoritism, because fewer neighbors share the same outgroup. This limits the scope of breakdowns in cooperation caused by outgroup homogeneity bias.} 
  \label{groups}
\end{figure}

Another parameter of interest is the number of groups, $m$.
This may be regarded as a measure of the population's diversity.
Increasing the number of groups does not affect ingroup cooperation, but increases outgroup cooperation, thus reducing ingroup favoritism (Figure~\ref{groups}).
Intuitively, if an individual's neighbors all belong to different groups, then tracking these groups' aggregate behavior is equivalent to tracking individual behavior.
The more diverse the population, the less meaningful outgroup homogeneity is as an approximation.
More practically, when fewer neighbors share group membership, breakdowns in cooperation result in smaller cascades of retributive defections.

\begin{figure}[t]
  \begin{center}
  \includegraphics[width=\linewidth]{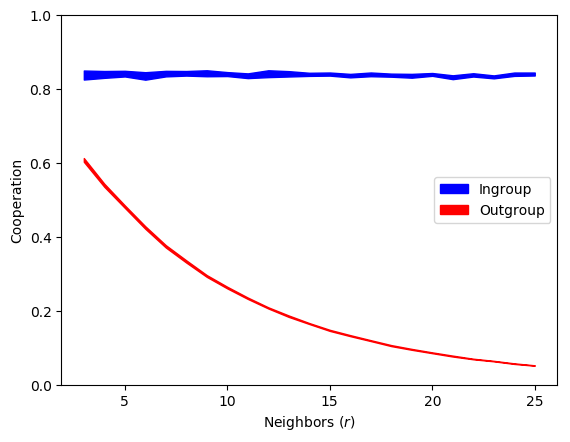}
  \end{center}
  \caption{Stabilized ingroup and outgroup cooperation rates for various neighborhood sizes ($r$). If there are many groups in the population (here $m=20$), reducing the number of neighbors reduces ingroup favoritism, by limiting the scope of defection cascades.}
  \label{degree}
\end{figure}

Finally, the number of neighbors that individuals interact with, $r$, is also relevant.
If there are relatively many groups in the population (e.g. $m=20$), then reducing the number of neighbors alleviates cascading breakdowns in cooperation, because fewer neighbors belong to the same group.
This promotes higher rates of outgroup cooperation, which in turn reduces ingroup favoritism (Figure~\ref{degree}).
However, if there are few groups (e.g. $m=2$), outgroup cooperation remains unsustainable, even if the number of neighbors is drastically reduced (e.g. to $r=3$), and ingroup favoritism remains high.

\section{Discussion}
We have presented an agent-based computational model of a PD game, where outgroup homogeneity causes ingroup favoritism between arbitrary groups.
Previous models have relied on indirect reciprocity (observing others' interactions) to produce such an outcome.
However, these only produced ingroup favoritism if they invoked additional factors.
For instance, \citeA{Masuda2012} found that reputation assignment rules had to differ for ingroup and outgroup members, while \citeA{Nakamura2012} found that the flow of reputation information had to be severed between groups.
By contrast, our model's results are driven by direct reciprocity (learning from personal experience), which obviates the need for additional assumptions about how others' interactions are evaluated, or how that information is shared.
The individuals we model leverage a minimal set of cognitive capacities.
Namely, they learn from past experience through Bayesian inference, and make rational decisions by maximizing expected utility.
In fact, they operate under similar assumptions to those found in game theoretical ``fictitious play"~\cite{Berger2007}:
They estimate others' probability of cooperating as a stationary strategy, and then select the best response to observed behavior.

In our model, outgroup homogeneity causes ingroup favoritism, because outgroup defections lower an entire group's perceived cooperativeness, rather than just the individual's.
Outgroup cooperation is difficult to establish and maintain not only because punishing a defector involves punishing its entire group, but also because it triggers a cascade of retributive action from those caught in the crossfire.
Our findings shed light on several empirical observations about ingroup favoritism.
For instance, ingroup favoritism is infamously easy to evoke even when it confers no advantage, and group membership is transparently arbitrary~\cite{Billig1973,Locksley1980}.
Moreover, meta-analysis suggests that ingroup favoritism between transient, experimentally-induced groups is often as strong as between natural ones~\cite{Balliet2014}.
The fact that this is often maladaptive from both the group and the individual's point of view makes it challenging to explain as a product of selection~\cite{Nakamura2012}.
In our model, individuals all share the same learning and decision rules, and are assigned groups at random.
There is no selection over genotypes.
Rather, strong ingroup favoritism arises rapidly as a phenotypic consequence of Bayesian learning, rational decision-making, and a well-established perceptual bias:
outgroup homogeneity.

One implication is that reducing or eliminating outgroup homogeneity bias may erode ingroup favoritism.
In reality, ingroup favoritism can be mitigated by taking the perspective of outgroup members~\cite{Galinsky2000}.
The apparent mechanism behind this result is that perspective-taking reduces reliance on group stereotypes~\cite{Brewer1996}, causing outgroup members to be perceived as individuals.
Greater intergroup contact may also reduce intergroup bias, both against that outgroup~\cite{Pettigrew2006}, as well as against uninvolved others~\cite{Pettigrew2009}.
In line with with our model's predictions, this too appears to be driven by reduced reliance on group stereotypes~\cite{Tadmor2012}.

Similarly, ever since Sherif's~\citeyear{Sherif1954} original Robbers Cave experiment, ingroup favoritism has often been both evoked and understood through the lens of competition~\cite{Sherif1961}.
In the experiment's final phase, intergroup tensions were deliberately reduced by encouraging cooperation.
Consistent with this view, our model predicts that incentivizing cooperation alleviates ingroup favoritism.
PD is a social dilemma precisely because it rewards both competition and cooperation.
Increasing the $b/c$ ratio thus minimizes these competitive aspects, and emphasizes the cooperative ones instead.
Reducing the temptation to defect, relative to the benefits of mutual cooperation, causes individuals to take more risks to establish mutual cooperation, and to recover it more readily when it breaks down.

Finally, our model also predicts that ingroup favoritism may be reduced by increasing population diversity.
When fewer neighbors belong to the same group, this limits the cascades of defection caused by outgroup homogeneity bias.
This is also why lowering the number of neighbors can be effective.
In both cases, the chances of being punished for an ingroup member's actions are reduced.
However, this reasoning only applies if group membership is indeed arbitrary.
The role of diversity in ingroup favoritism is typically studied through the lens of group differences, which add considerable complexity~\cite{Everett2015}.
Similarly, if conflicts exist along group lines, increased diversity may not necessarily reduce ingroup favoritism~\cite{Hewstone2014}.
No doubt, a great deal of real-world ingroup favoritism is intertwined with such pragmatic concerns.
However, because ingroup favoritism occurs even when such concerns are irrelevant, understanding such social factors could offer promising ways of addressing it.

\section{Acknowledgments}
We thank an anonymous reviewer for their helpful comments.

\bibliographystyle{apacite}

\setlength{\bibleftmargin}{.125in}
\setlength{\bibindent}{-\bibleftmargin}

\bibliography{references}

\end{document}